\documentclass[a4paper,UKenglish,cleveref, autoref, thm-restate]{lipics-v2021}

\pdfoutput=1 %
\hideLIPIcs  %

\graphicspath{{plots/}}%

\usepackage{xspace}
\usepackage{booktabs, multirow}
\usepackage{listings, listings-rust}
\usepackage{standalone}
\usepackage{tikz}

\definecolor{codegreen}{rgb}{0,0.6,0}
\definecolor{codegray}{rgb}{0.5,0.5,0.5}
\definecolor{codepurple}{rgb}{0.58,0,0.82}
\definecolor{backcolour}{rgb}{0.95,0.95,0.92}
\lstdefinestyle{mystyle}{
    backgroundcolor=\color{backcolour},   
    commentstyle=\color{codegreen},
    keywordstyle=\color{magenta},
    numberstyle=\tiny\color{codegray},
    stringstyle=\color{codepurple},
    basicstyle=\ttfamily\footnotesize,
    breakatwhitespace=false,         
    breaklines=true,                 
    captionpos=b,                    
    keepspaces=true,                 
    numbers=left,                    
    numbersep=5pt,                  
    showspaces=false,                
    showstringspaces=false,
    showtabs=false,                  
    tabsize=2
}
\newcommand\rustsat{\textsc{RustSAT}\xspace}
\newcommand\pysat{\textsc{PySAT}\xspace}
\newcommand{\pblib}{\textsc{PbLib}\xspace}
\newcommand{\openwbo}{\textsc{Open-WBO}\xspace}
\newcommand\cargo{\texttt{cargo}\xspace}

\def\rawrsversion{0.7.0}
\newcommand\doclink[1]{\texttt{#1}}
\newcommand\rsdocs[2]{\href{https://docs.rs/rustsat/\rawrsversion/rustsat/#2}{\doclink{#1}\xspace}}
\newcommand\rstrait[3]{\rsdocs{#1}{#2/trait.#3.html}}
\newcommand\crate[1]{\href{https://crates.io/crates/#1}{\doclink{#1}\xspace}}
\newcommand\module[1]{\rsdocs{#1}{#1}}

\newcommand\rsversion{\rawrsversion\xspace}

\title{\rustsat: A Library For SAT Solving in Rust} %

\author{Christoph Jabs}{University of Helsinki, Finland \and \url{https://christophjabs.info}}{christoph.jabs@helsinki.fi}{https://orcid.org/0000-0003-3532-696X}{}%

\authorrunning{Christoph Jabs} %

\Copyright{Christoph Jabs} %



\category{Tool paper} %

\relatedversion{} %

\supplement{\url{https://github.com/chrjabs/rustsat}}%

\funding{This work is funded by Research Council of Finland (grant 356046).}%

\acknowledgements{
Thanks to the people who have already contributed to \rustsat
on GitHub. Especially Christopher Jefferson, Jukka Suomela, and Noah Bruns have helped
greatly in providing feedback based on different usecases for
\rustsat.
I also want to thank Matti J\"arvisalo and Jeremias Berg, who have supported me
in working on \rustsat during my PhD research.
}%

\nolinenumbers %

\EventEditors{John Q. Open and Joan R. Access}
\EventNoEds{2}
\EventLongTitle{42nd Conference on Very Important Topics (CVIT 2016)}
\EventShortTitle{CVIT 2016}
\EventAcronym{CVIT}
\EventYear{2016}
\EventDate{December 24--27, 2016}
\EventLocation{Little Whinging, United Kingdom}
\EventLogo{}
\SeriesVolume{42}
\ArticleNo{23}

\begin{document}

\maketitle

\begin{abstract}
  State-of-the-art Boolean satisfiability (SAT) solvers constitute a
  practical and competitive approach for solving various real-world problems.
  To encourage their widespread adoption, the relatively high barrier of entry
  following from the low level syntax of SAT and the expert knowledge required
  to achieve
  tight integration with SAT solvers
  should be further reduced.
  We present \rustsat, a library with the aim of making SAT solving technology
  readily available in the Rust programming language.
  \rustsat provides functionality for helping with generating (Max)SAT
  instances, writing them to, or reading them from files.
  Furthermore, \rustsat includes interfaces to various state-of-the-art SAT solvers
  available with a unified Rust API.
  Lastly, \rustsat implements several encodings for higher level constraints
  (at-most-one, cardinality, and pseudo-Boolean), which are
  also available via a C and Python API.
\end{abstract}

\section{Introduction}

Boolean satisfiability (SAT) solving is a significant success story of recent
years~\cite{DBLP:journals/cacm/FichteBHS23}.
State-of-the-art SAT solvers are highly optimized reasoning engines, able to
solve propositional formulas with millions
of variables and clauses, as demonstrated, e.g., by recent SAT
Competitions~\cite{DBLP:journals/ai/FroleyksHIJS21, SatComp24}.
This has resulted in SAT solvers being commonly used as the underlying
reasoning engine for solving a wide variety of problems, ranging from model
checking~\cite{DBLP:reference/mc/BiereK18, DBLP:series/faia/Biere21} over
computer-assisted mathematics~\cite{DBLP:conf/sat/SubercaseauxH22,
DBLP:conf/tacas/HeuleS24, DBLP:conf/itp/SubercaseauxNGC24} to constraint
satisfaction solving~\cite{DBLP:conf/cp/PerronDG23}, and many more.
Through their commonly supported incremental solving
functionality~\cite{DBLP:journals/entcs/EenS03, DBLP:series/faia/0001LM21}, SAT
solvers also find application in problem-solving beyond
NP~\cite{DBLP:conf/spin/Clarke02, DBLP:series/faia/BarrettSST21}, and
in optimization~\cite{DBLP:series/faia/BacchusJM21}.

In order to achieve the level of performance optimization required,
state-of-the-art SAT solvers are written
in low-level languages, with C++ being the most common choice.
This can pose some challenges when building tools based on SAT
solving technology in different programming languages, as it often
requires writing additional ``boilerplate'' code in order to interface with SAT
solvers, which has to be repeated for each project and SAT solver.
The standardized IPASIR interface~\cite{ipasir} for incremental SAT solvers
removes some of the required work, by unifying the API for the most-basic
solver functionality.
For the Python ecosystem, the \pysat library~\cite{DBLP:conf/sat/IgnatievMM18,
DBLP:conf/sat/IgnatievTK24} has improved the accessibility of SAT solving
technology significantly by packaging interfaces to state-of-the-art SAT
solvers with helpful functionality for modelling problems as
propositional logic.
For interacting with SAT solvers from Python,
\textsc{OptiLog}~\cite{DBLP:conf/sat/AlosAST22} defines a standardized
\texttt{iSAT} C++ interface which includes more functionality than IPASIR.
However, as many languages (e.g., Rust) offer direct interaction only with C
APIs, \texttt{iSAT} is not universally applicable.

We present \rustsat, a library for
the Rust programming language, that includes functionality for building and
working with SAT instances,
interfaces to common state-of-the-art SAT solvers, and CNF encodings for
higher-level constraints.
\rustsat is geared towards performance-sensitive applications written in Rust, where either SAT
solvers are employed as part of the application, or propositional encodings for
problem settings are encoded and then written to file. 
Some applications where \rustsat is already being used are a product
configuration tool for the automotive
industry~\cite{Bruns2024ApplicationMultiObjective}, explaining pen and paper
puzzles~\cite{DBLP:journals/corr/abs-2104-15040} (originally based on \pysat,
but now using \rustsat), the \textsc{Scuttle} multi-objective MaxSAT
solver~\cite{scuttle}, all written in Rust, and the Loandra MaxSAT
solver~\cite{LoandraMSE24} using \rustsat through its C API.

\begin{figure}[t]
  \centering
  \includegraphics{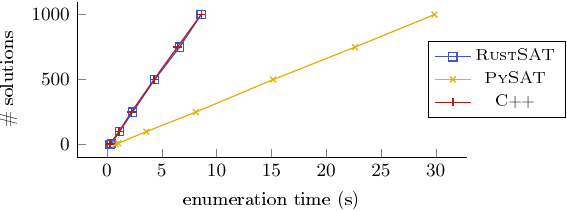}
  \caption{
  Runtime comparison of solution enumerator with \rustsat, \pysat, and C++ for the
  \href{https://benchmark-database.de/file/0f27eae382e7dcd8fd956ab914083a29?context=cnf}{\texttt{AProVE11-12.cnf}}
  instance.
  }\label{fig:enumeration}
\end{figure}

Being implemented in Rust, \rustsat provides a balance
between accessibility---through Rust's modern build system and package
distribution, as well as tooling to generate
convenient API documentation---and performance.
To illustrate the performance improvement that \rustsat can yield over
\pysat in certain applications, as well as the little overhead incurred
compared to not using any library at all, Figure~\ref{fig:enumeration} shows
the time
required to enumerate up-to 1000 solutions to the
\href{https://benchmark-database.de/file/0f27eae382e7dcd8fd956ab914083a29?context=cnf}{\texttt{AProVE11-12.cnf}}
instance from the SAT Competition 2022 Anniversary track, using \pysat,
\rustsat and a custom C++ implementation.%
\footnote{
Experiment run on an AMD Ryzen 7 1700 (3 GHz) with 16 GB of memory, using the
\texttt{pysat.examples.models} implementation included in \pysat (1.8.dev16)
and the \texttt{enumerator} tool from \crate{rustsat-tools}.
All three approaches use MiniSat 2.2.0 as the SAT solver.
}
While Python and \pysat can in many cases be
competitive~\cite{DBLP:conf/sat/IgnatievMM18, DBLP:journals/jsat/IgnatievMM19},
the more than \(3\times\)
speedup observed in Figure~\ref{fig:enumeration} shows the potential
efficiency improvement from implementing problem-solving tools using SAT
solvers in a more efficient programming language such as Rust.
Figure~\ref{fig:enumeration} also shows that \rustsat incurrs virtually no
overhead compared to implementing solution enumeration in C++, employing the
SAT solver directly.
Additionally, Rust provides the advantages of type and memory safety, detecting
many failure cases at compile time rather than at runtime.

All source code for \rustsat can be found at
\url{https://github.com/chrjabs/rustsat}, and detailed and up-to-date API
documentation is available at \url{https://docs.rs/rustsat}.
Throughout this paper, when we refer to Rust types, functions or crates,
we include hyperlinks to the API documentation online, formatted as
in this example: \rsdocs{rustsat}{}.

\section{The Rust Programming Language}\label{sec:rust}

We give a brief overview of those features of the Rust programming language and
its ecosystem that are relevant for understanding the design-principles of the \rustsat
library.

Rust~\cite{rust} is a programming language emphasizing
performance, type safety, and reliability.
In contrast to languages like Java or Python, Rust is fully compiled and does
not use a garbage collector for memory management.
In order to still achieve memory safety, Rust uses a system called the
\emph{borrow checker}, which during compilation tracks object lifetimes and
ensures that all memory references are valid.
For certain operations---e.g., when interfacing external C APIs or working with
raw pointers---the Rust compiler cannot ensure memory safety.
To enable usecases where such operations are necessary, Rust allows users to
write so-called \emph{unsafe} code, where memory safety guarantees need to
be manually upheld by the user, as e.g.~when programming in C++.

Code inside a Rust project is typically structured into several (sub)modules,
with paths to a type or function in a submodule being separated by two colons
(\texttt{::}), a notation we also use throughout this paper.
To enable generic code written against any type that supports specific
functionality, Rust supports abstraction via so-called \emph{traits}.
Traits work similarly to interfaces in, e.g., Java:
When defining a trait, one abstractly specifies certain methods that need to be
provided by implementors of the trait.
For example, a trait for SAT solvers might specify an \texttt{add\_clause} and a
\texttt{solve} method.
A type that implements the trait then supplies the concrete implementations of
the methods.
Code that is generically written against a trait rather than a specific type
can then be compiled with any type implementing that respective trait.

An important part of the Rust ecosystem is the \cargo package manager.
Similarly to the Python package manager \texttt{pip}, \cargo takes care of
downloading and compiling dependencies.
Additionally, \cargo provides a build system for projects, similar to
\texttt{cmake} or GNU Autotools for C++ projects.
A \cargo package is called a \emph{crate}, and crates
are typically published on \url{https://crates.io}.
The API documentation for crates on \texttt{crates.io} is automatically
generated and available at \url{https://docs.rs/}.

\section{Design Principles and Functionality}\label{sec:main}

\begin{figure}[t]
  \centering
  \begin{tikzpicture}[
  font=\small,
  crate/.style={rounded corners},
  mod/.style={rounded corners,dashed},
]
  \def\width{6}
  \def\height{3}

  \draw [crate] (0,0) rectangle (\width,\height);
  \node at (.5*\width,\height-.25) {\crate{rustsat}};

  \draw [mod] (.1,\height/2-.15) rectangle ++(\width/4-.15,\height/2-.35);
  \node at (\width/8+.0375,\height-.7) {\footnotesize\module{types}};

  \draw [mod] (.1,.1) rectangle ++(\width/4-.15,\height/2-.35);
  \node at (\width/8+.0375,\height/2-.45) {\footnotesize\module{instances}};

  \draw [mod] (\width/4+.05,\height/2-.15) rectangle ++(\width/2-.1,\height/2-.35);
  \node at (\width/2,\height-.7) {\footnotesize\module{solvers}};
  \node at (\width/2,\height-1) {\scriptsize Solver trait definitions};

  \draw [mod] (\width/4+.05,.1) rectangle ++(\width/2-.1,\height/2-.35);
  \node at (\width/2,\height/2-.45) {\footnotesize\module{encodings}};

  \node [mod,draw] at (\width/2-.9,\height/4-.2) {\footnotesize\rsdocs{am1}{encodings/am1}};
  \node [mod,draw] at (\width/2,\height/4-.2) {\footnotesize\rsdocs{card}{encodings/card}};
  \node [mod,draw] at (\width/2+.9,\height/4-.2) {\footnotesize\rsdocs{pb}{encodings/pb}};

  \draw [mod] (\width/4*3+.05,\height/2-.15) rectangle ++(\width/4-.15,\height/2-.35);
  \node at (\width/8*7-.0375,\height-.7) {\footnotesize\module{algs}};

  \node at (\width/8*7-.0375,\height/2-.45) {\footnotesize\(\dots\)};

  \draw [-latex,shorten <=3pt, shorten >=3pt] (-.6,\height-.7) -- (0,\height-.7);
  \node at (-\width/4-.6,\height-.25) {implement traits};

  \draw [crate] (-\width/2-.6,\height-.5) rectangle ++(\width/2,-.4);
  \node at (-\width/4-.6,\height-.7) {\crate{rustsat-kissat}};

  \draw [crate] (-\width/2-.6,\height-1) rectangle ++(\width/2,-.4);
  \node at (-\width/4-.6,\height-1.2) {\crate{rustsat-cadical}};

  \draw [crate] (-\width/2-.6,\height-1.5) rectangle ++(\width/2,-.4);
  \node at (-\width/4-.6,\height-1.7) {\crate{rustsat-minisat}};

  \draw [crate] (-\width/2-.6,\height-2) rectangle ++(\width/2,-.4);
  \node at (-\width/4-.6,\height-2.2) {\crate{rustsat-glucose}};

  \draw [crate] (-\width/2-.6,\height-2.5) rectangle ++(\width/2,-.4);
  \node at (-\width/4-.6,\height-2.7) {\crate{rustsat-batsat}};

  \draw [crate] (-\width/2-.6,\height-3) rectangle ++(\width/2,-.4);
  \node at (-\width/4-.6,\height-3.2) {\crate{rustsat-ipasir}};

  \draw [-latex,shorten <=3pt, shorten >=3pt] (\width+.6,\height-.7) -- (\width,\height-.7);
  \node at (\width/4*5+.6,\height-.25) {use Rust API};

  \draw [crate] (\width+.6,\height-.5) rectangle ++(\width/2,-.4);
  \node at (\width/4*5+.6,\height-.7) {C-API};

  \draw [crate] (\width+.6,\height-1) rectangle ++(\width/2,-.4);
  \node at (\width/4*5+.6,\height-1.2) {Python API};

  \draw [crate] (\width+.6,\height-1.5) rectangle ++(\width/2,-.4);
  \node at (\width/4*5+.6,\height-1.7) {\crate{rustsat-tools}};
\end{tikzpicture}
  \caption{The architecture of the \rustsat project. Crates are represented as
  solid boxes, and modules as dashed boxes.}\label{fig:architecture}
\end{figure}
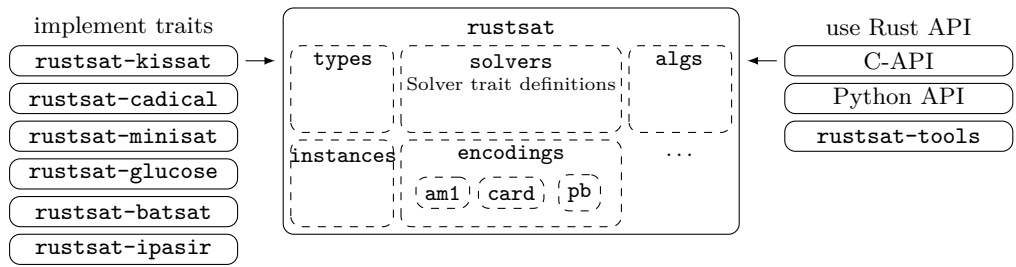

The architecture of \rustsat is illustrated in
Figure~\ref{fig:architecture}.
\rustsat consists of the main \crate{rustsat} crate, which contains most of
the functionality provided.
Interfaces to SAT solvers, as well as the C and Python APIs are provided as
separate crates.
Additionally, the \crate{rustsat-tools} crate contains some helpful
executable tools for working with SAT-related instances, e.g., a tool to verify
that a given assignment is indeed a solution to a CNF formula.

The aim of the \rustsat project is to be universal, convenient, and reliable to
use, while not sacrificing performance.
Towards this goal, we use an extensive automated test suite that is extended
whenever bugs are discovered, to avoid regressions as much as possible.
For reliability, we follow the semantic versioning scheme, where during this
quite early stage of development (marked by major version 0), breaking changes will
be published as a minor version bump, while any other changes are released as a patch
version.
All of \rustsat's functionality is accessible without requiring unsafe code.
The only unsafe elements in the \rustsat API are intended as slight performance
improvements when certain guarantees are already checked externally, and do
therefore not need to be verified by \rustsat (e.g.,
\rsdocs{Lit::new\_unchecked}{types/struct.Lit.html\#method.new\_unchecked}).
\rustsat is licensed under the MIT license, and developed entirely open-source.
User feedback and contributions have already been
very valuable in designing the API to be convenient in a wide range of
usecases.

In the following, we detail the main functionality included in \rustsat
version \rsversion split into functionality for working with SAT (and related) instances,
solver interfaces, and higher-level constraint encodings.

\subsection{Working with Problem Instances}

In the \rsdocs{types}{types} module, \rustsat defines various types that form
the basis of working with SAT instances in Rust, and of the typesafe API of
\rustsat.
The type \rsdocs{Var}{types/struct.Var.html} wraps a 32-bit integer to
represent propositional variables (starting from index 0), and
\rsdocs{Lit}{types/struct.Lit.html} represents literals as the variable index
shifted one bit to the left, and the last bit signaling negation, i.e., with
the same memory representation as MiniSat~\cite{DBLP:conf/sat/EenS03}.
The methods \rsdocs{Lit::from\_ipasir}{types/struct.Lit.html\#method.from\_ipasir}
and \rsdocs{Lit::to\_ipasir}{types/struct.Lit.html\#method.to\_ipasir} convert
literals to the integer representation used in IPASIR~\cite{ipasir} and the
DIMACS CNF~\cite{dimacs-cnf} format.
The \rsdocs{Clause}{types/constraints/struct.Clause.html} type is a wrapper around a
vector providing some helpful functionality for working with clauses.
Whenever a clause is referenced, \rustsat accepts any slice of literals (i.e., contiguous sequence
in memory), making the \rsdocs{Clause}{types/constraints/struct.Clause.html} type optional to use.
To build instances of satisfiability or optimization problems, the
\rsdocs{SatInstance}{instances/sat/struct.Instance.html} and
\rsdocs{OptInstance}{instances/opt/struct.Instances.html} types are available
in the \rsdocs{instances}{instances} module.%
\footnote{If compiled with an additional feature flag, a
\rsdocs{MultiOptInstance}{instances/multiopt/struct.Instances.html} type is
available for multi-objective optimization.}
As their constraints, these instance types support clauses, cardinality,
and pseudo-Boolean constraints.
To convert them to conjunctive normal form
(\rsdocs{Cnf}{instances/sat/struct.Cnf.html}), the constraint encodings
described in Section~\ref{sec:encodings} can be used.
Objectives (\rsdocs{Objective}{instances/opt/struct.Objective.html}) can be
represented either on the basis of soft clauses or as a linear function over
literals.
Instance objects can also be crated from DIMACS CNF~\cite{dimacs-cnf} and
WCNF~\cite{dimacs-wcnf}, as well as OPB~\cite{opb} files via parser implementations
included in \rustsat.
As an illustration of the API, slightly simplified method signatures of
the \rsdocs{SatInstance}{instances/sat/struct.Instance.html} type are shown in
Listing~\ref{list:inst}.

\begin{lstlisting}[
  float=t,
  caption={Excerpt of the \rsdocs{SatInstance}{instances/sat/struct.Instance.html} API},
  label={list:inst},
]
impl SatInstance {
  // Adding constraints
  pub fn add_clause(&mut self, cl: Clause);
  pub fn add_card_constr(&mut self, card: CardConstraint);
  pub fn add_pb_constr(&mut self, pb PbConstraint);
  pub fn add_lit_impl_clause(&mut self, a: Lit, b: &[Lit]);
  // File parsing
  pub fn from_dimacs_path(path: &Path) -> Result<Self>;
  pub fn from_opb_path(path: &Path, opts: opb::Options) -> Result<Self>;
  // Converting to CNF
  pub fn into_cnf(self) -> (Cnf, VarManager);
  // ...
}
\end{lstlisting}

Variable management in \rustsat is handled by so-called variable managers that
implement the \rsdocs{ManageVars}{instances/trait.ManageVars.html} trait.
Most central to the \rsdocs{ManageVars}{instances/trait.ManageVars.html} trait
is the \rsdocs{new\_var}{instances/trait.ManageVars.html\#tymethod.new\_var}
method, which is used for obtaining a variable that has not been used so far.
The simplest variable manager simply keeps track of the next free variable
index (\rsdocs{BasicVarManager}{instances/struct.BasicVarManager.html}), while
more advanced variable managers can maintain a mapping between propositional
variables and any Rust objects
(\rsdocs{ObjectVarManager}{instances/struct.ObjectVarManager.html}).

\subsection{SAT Solver Interfaces}

The vast majority of state-of-the-art SAT solvers offer APIs in either C/C++.
While calling C APIs from Rust is possible, the APIs for solvers often differ,
with the IPASIR~\cite{ipasir} interface for incremental solver standardizing only
the most-central functionality.
Furthermore, tightly integrating the build system of each solver with Rust's
\cargo build system to make them accessible in Rust requires additional work.
\rustsat standardizes interfaces for common functionality in SAT solvers via a
series of traits that are defined in the \rsdocs{solvers}{solvers} module of
the main \crate{rustsat} crate.
In separate crates, \rustsat packages various state-of-the-art SAT solvers that
can be easily integrated in the \cargo build system.
These solver interface crates implement the traits defined in the \crate{rustsat} crate,
which enables easily swapping out a given SAT solver for another one that
provides the same required functionality.

We first describe the traits defined by \rustsat, which unify the solver interfaces.
Afterwards, we list the solvers \rustsat currently provides interfaces for,
and which functionality they support.

\newcommand\solvertrait[1]{\rstrait{#1}{solvers}{#1}}
All traits capturing SAT solver functionality are defined and documented in the
\rsdocs{solvers}{solvers} module.
The documentation of this module also gives an up-to-date overview of the SAT
solvers that are supported.
The most basic functionality that each SAT solver supports is captured in the
\solvertrait{Solve} trait.
This trait includes methods for adding clauses, solving without assumptions,
and obtaining solutions.
\emph{Incremental} solving~\cite{DBLP:journals/entcs/EenS03,
DBLP:series/faia/0001LM21} is supported via the \solvertrait{SolveIncremental}
trait, which contains the
\rsdocs{solve\_assumps}{solvers/trait.SolveIncremental.html\#tymethod.solve\_assumps}
and \rsdocs{core}{solvers/trait.SolveIncremental.html\#tymethod.core} methods
for solving under assumptions and obtaining a core.
An overview of the most important methods in the \solvertrait{Solve} and
\solvertrait{SolveIncremental} traits is given in Listing~\ref{list:solve}.
Additional traits provide functionality for terminating solvers via callbacks
(\solvertrait{Terminate}, compare \texttt{ipasir\_set\_terminate}~\cite{ipasir}) or
an asynchronous interrupt signal
(\solvertrait{Interrupt}), tracking learned clauses (\solvertrait{Learn}, compare
\texttt{ipasir\_set\_learn}~\cite{ipasir}), setting preferred phases for variables
(\solvertrait{PhaseLit}), excluding variables from being eliminated by
pre/inprocessing (\solvertrait{FreezeVar}), flipping literals in found solutions
(\solvertrait{FlipLit}), propagating assumptions (\solvertrait{Propagate}),
setting limits (\solvertrait{LimitConflicts}, \solvertrait{LimitDecisions},
\solvertrait{LimitPropagations}), and getting solver statistis
(\solvertrait{SolveStats}, \solvertrait{GetInternalStats}).

\begin{lstlisting}[
  float=t,
  caption={Excerpt of the \solvertrait{Solve} and \solvertrait{SolveIncremental} traits.},
  label={list:solve},
]
pub trait Solve {
  fn signature(&self) -> &'static str;
  fn solve(&mut self) -> Result<SolverResult>;
  fn lit_val(&mut self, lit: Lit) -> Result<TernaryVal>;
  fn add_clause(&mut self, clause: Clause) -> Result<()>;
  // ...
}

pub trait SolveIncremental {
  fn solve_assumps(&mut self, assumps: &[Lit]) -> Result<SolverResult>;
  fn core(&mut self) -> Result<Vec<Lit>>;
}
\end{lstlisting}

\rustsat provides crates for the state-of-the-art SAT solvers
Kissat~\cite{BiereFallerFazekasFleuryFroleyksPollitt-SAT-Competition-2024-solvers},
CaDiCaL~\cite{DBLP:conf/cav/BiereFFFFP24}, MiniSat~\cite{DBLP:conf/sat/EenS03},
and Glucose~\cite{DBLP:conf/ijcai/AudemardS09}.
The detailed solver versions can be seen in Table~\ref{tab:solvers}; for some
solvers multiple versions are supported, with the used version being selected
via feature flags at compilation time.
All except the first are incremental solvers and therefore implement the
\rsdocs{SolveIncremental}{solvers/trait.SolveIncremental.html} trait.
CaDiCaL supports the most functionality, implementing all mentioned traits
except for \solvertrait{LimitPropagations}.
Details on which additional traits each solver implements can be found in the online
documentation of the solver crates.
In addition to those well-known solvers, \rustsat also implements an interface
to BatSat~\cite{batsat}, which is a reimplementation
of MiniSat in Rust.
Choosing BatSat allows for projects to be pure Rust, and therefore compile to,
e.g., web assembly.

\begin{table}[t]
  \caption{SAT solver interfaces available for \rustsat.}\label{tab:solvers}
  \centering
  \begin{tabular}{@{}lllr@{}}
    \toprule
    \multicolumn{1}{c}{Solver} & \multicolumn{1}{c}{Versions} & \multicolumn{1}{c}{Crate} & \multicolumn{1}{c}{Incremental} \\
    \midrule
    Kissat~\cite{BiereFallerFazekasFleuryFroleyksPollitt-SAT-Competition-2024-solvers} & 3.0.0 -- 4.0.2 & \crate{rustsat-kissat} & no \\
    CaDiCaL~\cite{DBLP:conf/cav/BiereFFFFP24} & 1.5.0 -- 2.1.3 & \crate{rustsat-cadical} & yes \\
    MiniSat~\cite{DBLP:conf/sat/EenS03} & 2.2.0 & \crate{rustsat-minisat} & yes \\
    Glucose~\cite{DBLP:conf/ijcai/AudemardS09} & 4.2.1 & \crate{rustsat-glucose} & yes \\
    Batsat~\cite{batsat} & 0.6.0 & \crate{rustsat-batsat} & yes \\
    IPASIR & -- & \crate{rustsat-ipasir} & yes \\
    Call solver binary & -- & \rsdocs{ExternalSolver}{solvers/external/struct.Solver.html} & no \\
    \bottomrule
  \end{tabular}
\end{table}

In addition to solver interfaces usable ``out of the box'', via the \crate{rustsat-ipasir} crate \rustsat
allows for linking to any user-provided library which implements IPASIR~\cite{ipasir}.
While CaDiCaL does implement the IPASIR interface, using CaDiCaL through its
dedicated interface gives access to more functionality: the
\crate{rustsat-ipasir} crate only implements the \solvertrait{Solve},
\solvertrait{SolveIncremental}, \solvertrait{Learn}, \solvertrait{Term}, and
\solvertrait{SolveStats} traits.
Lastly, the \rsdocs{ExternalSolver}{solvers/external/struct.Solver.html}
type allows for calling an executable solver binary with DIMACS input, and
parsing the output written to the standard output interface, given that it
follows the output specification of the SAT competition~\cite{sat-comp-output}.
This can be convenient for using solvers such as
Gimsatul~\cite{DBLP:journals/corr/abs-2207-13577} that are not intended to be used as
libraries.

\subsection{Constraint Encodings}\label{sec:encodings}

In the \rsdocs{encodings}{encodings} module, \rustsat implements CNF
encodings for higher-level constraints.
Beyond simple encodings (e.g., a literal implying a clause,
\rsdocs{atomics::lit\_impl\_clause}{encodings/atomics/fn.lit\_impl\_clause.html}),
mainly intended to
increase code readability, encodings for
at-most-one, cardinality, and pseudo-Boolean constraints are available.
Providing efficiently implemented constraint encodings in \rustsat allows users
to employ state-of-the-art constraint encodings when solving real-world
problems without having to go through the error-prone process of implementing
complex encodings themselves.
Interfaces to the different constraint encodings are unified via traits: for
at-most-one constraint encodings, the
\rsdocs{am1::Encode}{encodings/am1/trait.Encode.html} trait captures all
functionality, while for cardinality and pseudo-Boolean constraint encodings
upper and lower bounding, as well as incremental and non-incremental use are
split into separate traits in the \rsdocs{card}{encodings/card} and
\rsdocs{pb}{encodings/pb} submodules.
With incremental use of encodings we refer to either adding additional input
literals, or changing the enforced bound, both while reusing previously built
parts of the encoding~\cite{DBLP:conf/cp/MartinsJML14}.
Table~\ref{tab:encodings} lists the encodings that are currently implemented
and which functionality they support.
To the best of our knowledge, \rustsat is currently the only readily available
constraint encoding library providing an implementation of the dynamic
polynomial watchdog encoding~\cite{DBLP:conf/sat/PaxianR018}, which is used in
state-of-the-art MaxSAT solvers~\cite{PacoseMSE24}.

\begin{table}[t]
  \caption{The constraint encodings implemented in \rustsat. Bound types
  supported by a specific encoding are indicated as upper bounds (\(\le\)) and
  lower bounds (\(\ge\)).}\label{tab:encodings}
  \centering
  \begin{tabular}{@{}llrr@{}}
    \toprule
    \multicolumn{1}{c}{Constraint} & \multicolumn{1}{c}{Encoding} & \multicolumn{1}{c}{Bounds} & \multicolumn{1}{c}{Incrementality} \\
    \midrule
    Pseudo-Boolean & \rsdocs{GeneralizedTotalizer}{encodings/pb/gte/struct.GeneralizedTotalizer.html}~\cite{DBLP:conf/cp/0001MM15} & \(\le\) & yes \\
                   & \rsdocs{BinaryAdder}{encodings/pb/adder/struct.BinaryAdder.html}~\cite{DBLP:journals/ipl/Warners98,DBLP:journals/jsat/EenS06} & \(\ge, \le\) & yes \\
                   & \rsdocs{DynamicPolyWatchdog}{encodings/pb/dpw/struct.DynamicPolyWatchdog.html}~\cite{DBLP:conf/sat/PaxianR018} & \(\le\) & only changing bounds \\
    \midrule
    Cardinality & \rsdocs{Totalizer}{encodings/card/totalizer/struct.Totalizer.html}~\cite{DBLP:conf/cp/BailleuxB03,DBLP:conf/cp/MartinsJML14} & \(\ge, \le\) & yes \\
    \midrule
    At-most-one & \rsdocs{Pairwise}{encodings/am1/struct.Pairwise.html}~\cite{DBLP:series/faia/Prestwich21} & \(\le 1\) & no \\
                & \rsdocs{Ladder}{encodings/am1/struct.Ladder.html}~\cite{gent2004new} & \(\le 1\) & no \\
                & \rsdocs{Bitwise}{encodings/am1/struct.Bitwise.html}~\cite{doi:https://doi.org/10.1002/9780470612309.ch15} & \(\le 1\) & no \\
                & \rsdocs{Commander}{encodings/am1/struct.Commander.html}~\cite{Klieber2007EfficientCE} & \(\le 1\) & no \\
                & \rsdocs{Bimander}{encodings/am1/struct.Bimander.html}~\cite{DBLP:conf/soict/NguyenM15}  & \(\le 1\) & no \\
    \bottomrule
  \end{tabular}
\end{table}

For all constraint encodings implemented in \rustsat, we aim to produce the
smallest number of clauses required to enforce the constraint that is being
encoded.
Towards this end, we employ the \emph{cone of influence}
strategy~\cite{DBLP:conf/sat/PaxianR018} which removes clauses from the
encoding which contain pure literals---i.e., literals that only appear in one
polarity---either directly, or after having already removed other clauses.
Removing these clauses preserves both the correctness and the propagation
properties of the encoding.

\begin{figure}[b!]
  \centering
  \begin{tabular}{ccl}
    \hspace{1.5em}Totalizer & Binary adder & \hspace{1em}Generalized Totalizer \\
    \includegraphics{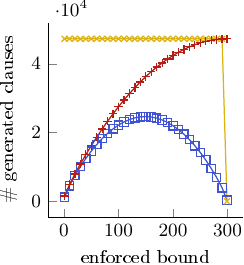} &
    \includegraphics{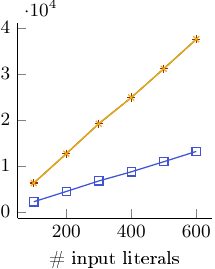} &
    \includegraphics{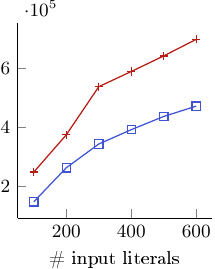} \\
    \multicolumn{3}{c}{\includegraphics{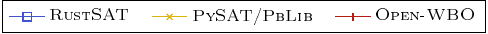}} \\
  \end{tabular}
  \caption{
  Left: Number of clauses in a totalizer cardinality encoding over 300 input
  literals for a given bound.
  Middle/Right: Number of clauses for a binary adder (middle) and generalized totalizer
  (right) pseudo-Boolean encoding for a given number of input literals with random weight
  in \([1,100]\) and enforced bound of 300.
  }\label{fig:coi}
\end{figure}

To illustrate the effect that the cone of influence strategy has, in
Figure~\ref{fig:coi}, we compare the number of clauses in CNF encodings
generated by \rustsat (\rsversion), \pysat (1.8.dev16, note that the
pseudo-Boolean constraint encodings provided are reexported from
\pblib~\cite{DBLP:conf/sat/PhilippS15} via the \textsc{PyPBLib} interface), and \openwbo
(2.1)~\cite{DBLP:conf/sat/MartinsML14}.
For the totalizer cardinality encoding~\cite{DBLP:conf/cp/BailleuxB03},
Figure~\ref{fig:coi} (left) shows the number of clauses in a totalizer
encoding produced for 300 input literals and an enforced upper bound ranging
from 1 to 300.
It can be observed that \pysat seems to be generating the clauses required for
enforcing \emph{any} bound at all times.
\openwbo omits clauses only required to define the semantics for output
variables corresponding to higher values than the currently enforced bound.
In \rustsat, we also omit encoding output variables that are lower than the
required range, which results in smaller encodings for higher bounds.
Figure~\ref{fig:coi} also shows the number of generated clauses for binary
adder (middle) and generalized totalizer (right) pseudo-Boolean encodings generated for
100--600 input literals with random weights in \([1,100]\) and an enforced bound
of 300.
Also for the pseudo-Boolean encodings, \rustsat produces the fewest clauses,
with \pblib and \openwbo producing both more than \(3\times\) the number of
clauses for the binary adder, while the saving for the generalized totalizer
encoding is smaller.
Note that neither \pblib nor \pysat implement the generalized totalizer encoding.

In addition to the encodings listed in Table~\ref{tab:encodings}, the
\rsdocs{card::simulators}{encodings/card/simulators} and
\rsdocs{pb::simulators}{encodings/pb/simulators} submodules contain helpers for
inverting encodings (i.e., encoding an upper bound constraint with a
lower-bounding encoding by inverting the constraint, and vice versa;
\rsdocs{Inverted}{encodings/card/simulators/struct.Inverted.html}), combining
an upper and a lower-bounding encoding to get an encoding for both bound
types at the same time, (\rsdocs{Double}{encodings/card/simulators/struct.Double.html}) and
encoding a pseudo-Boolean constraint by expanding it into a cardinality
constraint with repeated input literals and using a cardinality encoding
(\rsdocs{Card}{encodings/pb/simulators/struct.Card.html}).

Since the recent 0.7.0 release of \rustsat, the
\rsdocs{Totalizer}{encodings/card/totalizer/struct.Totalizer.html} and
\rsdocs{GeneralizedTotalizer}{encodings/pb/gte/struct.GeneralizedTotalizer.html}
encodings provide functionality for certifying the correctness of the generated CNF encoding
by producing a proof in \textsc{VeriPB} format~\cite{VeripbSC23, DBLP:conf/lpnmr/VandesandeWB22, DBLP:conf/tacas/JabsBBJ25}.
With this feature, the encodings can be employed in certified MaxSAT solvers,
or for building a tool that produces certified translations from OPB to CNF.

\subsection{C and Python API}

While the primary intended way of using \rustsat is from the Rust programming
language, some of its functionality is also exposed via a C and a Python API.

In version \rsversion, the C API contains access for all higher-level
constraint encodings listed in Table~\ref{tab:encodings}.
In the C API, literals are represented as IPASIR-style \texttt{int}s, and
clauses are returned via callbacks that work similarly to
\texttt{ipasir\_add}~\cite{ipasir}.
The main usecase for the C API is using the (incremental) encoding
implementations in solvers written in a different language, as is for example the case
for the Loandra MaxSAT solver~\cite{LoandraMSE24}.
To use the \rustsat C API in a project, the
\href{https://github.com/chrjabs/rustsat/blob/main/capi}{\doclink{rustsat-capi}}
crate needs to be compiled, which produces a statically linkable library.
The full documentation of the C API can be found in
\href{https://github.com/chrjabs/rustsat/blob/main/capi/rustsat.h}{\doclink{rustsat.h}}.

Similarly to the C API, the Python API exposes all constraint encodings
included in \rustsat.
Additionally, the \rsdocs{Lit}{types/struct.Lit.html} and
\rsdocs{Cnf}{instances/sat/struct.Cnf.html} types, as well as a variable manager are
included.
At this point, the Python API of \rustsat does not include solver interfaces,
but the \rustsat encodings can be used together with the solver interfaces in
\pysat.
The Python API is published on PyPI (\url{https://pypi.org/project/rustsat})
and its documentation is available online
(\url{https://christophjabs.info/rustsat/pyapi}).

\section{Conclusions}\label{sec:conclusions}

We presented \rustsat version \rsversion, a library with the aim of making SAT
solving technology more accessible from the Rust programming language.
\rustsat packages tools for dealing with satisfiability and optimization
instances in Rust, unified interfaces to state-of-the-art SAT solvers, and CNF
encodings for at-most-one, cardinality, and pseudo-Boolean constraints.
We have given an overview of the design principles of \rustsat---providing an
easy-to-use API while not sacrificing performance---and illustrated key
features empirically.
\rustsat is available in open source, and detailed
and up-to-date API documentation can be found online.

\bibliography{paper,dblp}

\appendix

\end{document}